\documentclass[iop]{emulateapj}
 \usepackage{psfrag}
 \usepackage{rotating}
 
\shorttitle{Limb-darkening \& Planetary Transits}
\shortauthors{Neilson~et~al.}

\begin{document}

\title{Limb Darkening and Planetary Transits: Testing Center-to-limb Intensity Variations and Limb-Darkening Directly from Model Stellar Atmospheres}

 \author{Hilding R.~Neilson\altaffilmark{1}}
 \altaffiltext{1}{Department of Astronomy \& Astrophysics, University of Toronto, 50 St.~George Street, Toronto, ON, M5S 3H4, Canada}
 \altaffiltext{2}{Department of Physics \& Astronomy, East Tennessee State University, Box 70652, Johnson City, TN 37614 USA}
\email{neilson@astro.utoronto.ca}

\author{Joseph T.~McNeil\altaffilmark{2, 3}} 
\altaffiltext{3}{Department of Physics, University of Florida, P.O.~Box 118440, Gainesville, FL 32611}
\author{Richard Ignace \altaffilmark{2}}
\author {John B. Lester\altaffilmark{4,1}}
\altaffiltext{4}{Department of Chemical \& Physical Sciences, University of Toronto Mississauga, Mississauga, ON L5L 1C6, Canada}

\begin{abstract}
The transit method, employed by MOST, \emph{Kepler}, and various ground-based surveys has enabled  the characterization of extrasolar planets to unprecedented precision.  These results are precise enough to begin to measure planet atmosphere composition, planetary oblateness, star spots, and other phenomena at the level of a few hundred parts-per-million.  However, these results depend on our understanding of stellar limb darkening, that is, the intensity distribution across the stellar disk that is sequentially blocked as the planet transits.  Typically, stellar limb darkening is assumed to be a simple parameterization with two coefficients that are derived from stellar atmosphere models or fit directly.  In this work, we revisit this assumption and compute synthetic planetary transit light curves directly from model stellar atmosphere center-to-limb intensity variations (CLIV) using the plane-parallel \textsc{Atlas} and spherically symmetric \textsc{SAtlas} codes. We compare these light curves to those constructed using best-fit limb-darkening parameterizations.  We find that adopting parametric stellar limb-darkening laws lead to systematic differences from the more geometrically realistic model stellar atmosphere CLIV of about 50 -- 100 ppm at the transit center and up to 300 ppm at ingress/egress.  While these errors are small they are systematic, and appear to limit the precision necessary to measure secondary effects. Our results may also have a significant impact on transit spectra.
\end{abstract}

\section{Introduction}
The discovery of extrasolar planets in the past two decades has revolutionized our view of the universe and the prospects of discovering life around other stars.  The rate of discovery has grown exponentially thanks to the transit detection method \citep{Charbonneau2004}  implemented through surveys such as the \textit{Kepler} satellite \citep{Borucki2010,Koch2010} and WASP \citep{Pollacco2006}.  The results from these surveys are finding an assortment of planets ranging from super Earths to hot Jupiters, challenging the traditional picture of planet formation and evolution.  However, to understand these transiting planets it is also necessary to understand their host stars to achieve important constraints on the size of the planets and
whether they have atmospheres.

\cite{Mandel2002} developed an analytic method that is commonly employed for fitting transit light curves that requires understanding the star's radius and center-to-limb intensity variation (CLIV).  As the planet passes in front of the star it blocks a small fraction of the star's light, hence tracing a chord of the center-to-limb intensity variation.  Typically, the intensity is represented by a simple limb-darkening law with either two or four free parameters \citep{Claret2000}, similar to the analysis for understanding eclipsing binary light curves. However, because both stars in the eclipsing binary are bright, the 
light curves do not constrain the CLIV well \citep[\textit{e.g.}][]{Popper1984}.  Even though observations of eclipsing binary light curves have improved dramatically, the light curves are still fit using  simple formulations for limb-darkening \citep{Prsa2005, Kirk2016}.  Similarly, microlensing observations provide even weaker constraints on limb-darkening laws \citep{Dominik2004, Fouque2011}.   On the other hand, interferometric observations are beginning to probe the details of stellar limb darkening, such as the impact of convection \citep{Chiavassa2010}, and the importance of extension in cool massive stars \citep[e.g.,][]{Wittkowski2004, Wittkowski2006a, Wittkowski2006b}.

Similarly, the precision of planetary transit observations is now 
approaching the point where the details of the stellar atmosphere CLIV
are becoming important for the analysis of the planet.  To that end, \cite{Sing2009} and \cite{Sing2010} developed a three-parameter limb-darkening law that is a significant improvement on the quadratic law, but provides no improvement to the \cite{Claret2000} four-parameter law. Furthermore,  \cite{Espinoza2015, Espinoza2016} carefully analyzed the biases induced by assuming various limb-darkening laws and found that the three-parameter law along with other square-root and logarithmic limb-darkening laws can provide more accurate measurements of transit properties than more commonly used linear and quadratic limb-darkening laws.

  In turn, the transit light curves can also be used to test model stellar atmospheres. For example, \cite{Knutson2007} fit multi-wavelength observations of HD~209458 to measure limb-darkening coefficients that were  compared with synthetic limb-darkening coefficients computed using the \textsc{Atlas} stellar atmosphere code \citep{Kurucz1979}.  However, theory and observations disagreed.  \citet{Claret2009} also analyzed the observations of HD~209458 using 
limb-darkening coefficients computed with model stellar atmospheres that
included different turbulent velocities, different procedures to compute the 
limb-darkening coefficients and even different atmosphere codes, but  disagreements between the predicted and observed light 
curves remained. This suggests a tension between theory and observations that needs to be resolved.

HD~209458 is not the only system challenging our understanding of limb darkening and stellar atmospheres. For instance, disagreements have been found to exist between theory and observations for the systems XO-1b, HAT-P-1b \citep{Winn2007}, HAT-P-11b \citep{Deming2011}, Kepler-5b \citep{Kipping2011a, Kipping2011b}, CoRot-13b \citep{Cabrera2010} and others.  These disagreements, while not greatly affecting the measurements of  the planetary properties, do raise questions about model stellar atmospheres and the precision of the light-curve fits. 

\cite{Howarth2011} found that, at least for some cases, the disagreements can be resolved by accounting for the inclination of the system, defined by the impact parameter,  $b$, the minimum distance between the center of the stellar disk
and the transiting planet's path normalized to the star's radius,  
\begin{equation} \label{eq:def_impact_par} 
b \equiv \frac{a\cos i}{R_*},
\end{equation}
where $a$ is the planet's orbital semimajor axis and $i$ is the orbital inclination.  For a planet with an orbit inclined to our line of sight, the transit path is only a limited chord
across the stellar disk, not the full diameter. Therefore, the limb-darkening coefficients must be altered 
accordingly, especially if there are errors in fitting the CLIV.  Similar results were found by \cite{Muller2013}.
\begin{figure*}[th]
\begin{center}
\plottwo{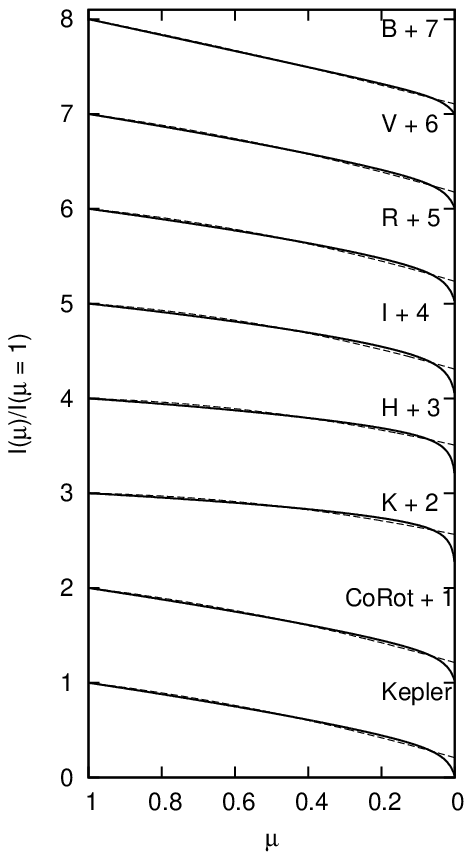}{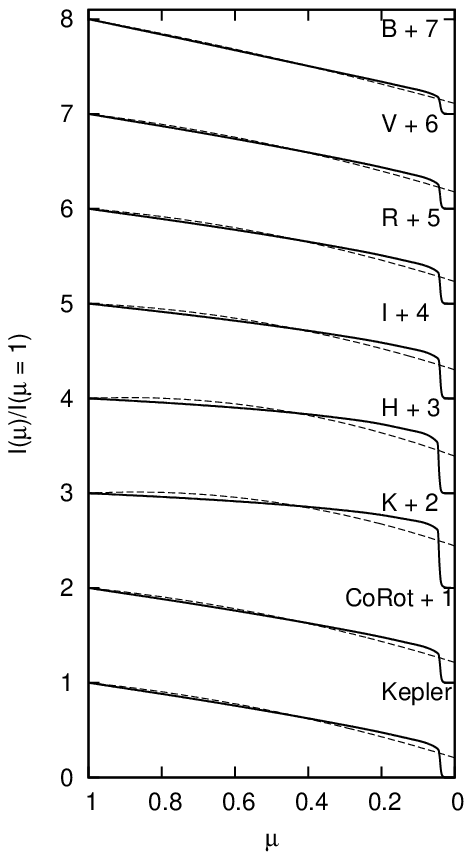}
\end{center}
\caption{Comparison of wavelength-dependent model stellar atmosphere CLIV (solid line) with the corresponding best-fit quadratic limb-darkening law (dashed line) as a function of $\mu \equiv \cos \theta$, where, for clarity, each waveband's 
profile is offset by the number next to the band's name. (left) 
Plane-parallel model atmosphere with $T_\mathrm{eff} = 5600$~K and 
$\log g = 4.5$. (right) Spherically symmetric geometry with luminosity $L = 0.8~L_\odot$, radius $R = 1.0~R_\odot$ and mass $M = 1.1~M_\odot$.}
\label{f1}
\end{figure*}

While the results of \citet{Howarth2011} and \citet{Muller2013} proved 
promising, they did not resolve the differences for all systems, and other
factors have been considered.  For example, star spots will change the stellar flux and the apparent effective
temperature, meaning the theoretical limb-darkening coefficients will be
computed from the wrong stellar atmosphere model \citep{Csizmadia2013}.
Another factor is the method for fitting the limb-darkening 
coefficients.  \citet{Kipping2013} showed that observational fits are 
more sensitive to a linear combination of the limb-darkening coefficients that 
makes them more linearly dependent.  
A third factor is  the role of the geometry of the stellar atmosphere on 
the star's CLIV; spherically symmetric 
model atmospheres produce different limb-darkening coefficients than the
more commonly used plane-parallel model stellar atmospheres
\citep{Neilson2011,Neilson2012,Neilson2013a,Neilson2013b}.  However, none of these 
solutions have been shown to resolve all the differences.

In this work, we compare planetary transit light curves computed using limb-darkening laws with those calculated directly from the model stellar atmosphere CLIV.  We start in Section~\ref{s2} by outlining the analytic approach for planet transits \citep{Mandel2002} to better understand the role of limb darkening .  In Section~\ref{s3}, we describe the model atmosphere code and models chosen for this work as well as how the model limb-darkening coefficients are computed.  In Section~\ref{s4}, we present results of the comparison between light curves computed using model CLIV and corresponding limb-darkening coefficients from a traditional two-parameter quadratic limb-darkening law.  We compare the CLIV predictions with light curves computed assuming a more complex four-parameter limb-darkening law in Section~\ref{s5}. In Section~\ref{s6}, we extend the analysis to consider the sensitivity of limb darkening with respect to the impact parameter, \textit{i.e.}, different orbital inclinations, and in Section~\ref{s7} we model the differences as a function of wavelength for transit spectroscopy. We summarize our work in Section~\ref{s8}.
\begin{figure*}[th]
\begin{center}
\plottwo{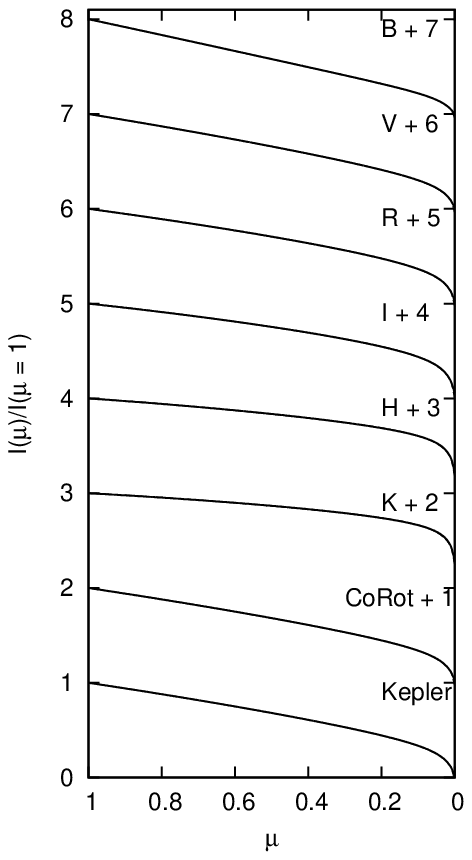}{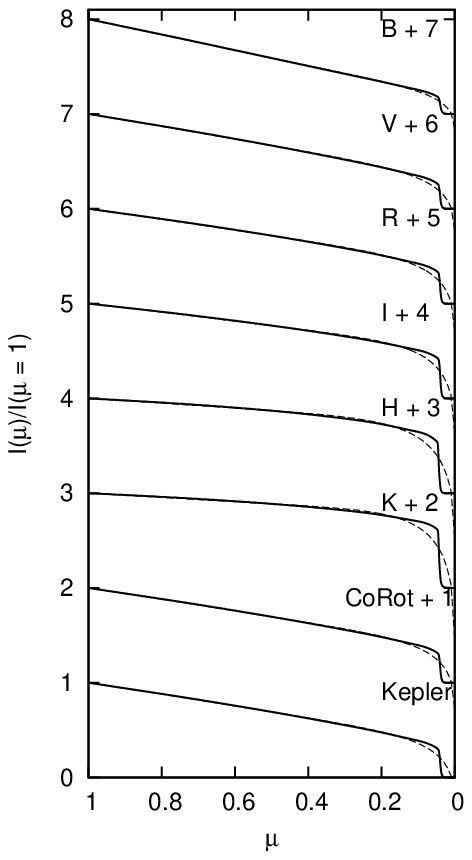}
\end{center}
\caption{Comparison of wavelength-dependent model stellar atmosphere CLIV (solid line) with corresponding best-fit \cite{Claret2000} four-parameter limb-darkening law (dashed line) as a function of $\mu \equiv \cos \theta$, where the model atmosphere assumes plane-parallel (left) and spherically symmetric geometry (right). The best-fit limb-darkening laws are indistinguishable from the plane-parallel CLIV in the left panel.}
\label{f2}
\end{figure*}

\section{Method}\label{s2}
We compute planetary transit light curves using either model stellar atmosphere CLIV or limb-darkening laws.  We use model stellar atmosphere CLIV from 
\citet{Neilson2013b}, computed using versions of the 
\textsc{Atlas9} stellar atmospheres code \citep{Kurucz1979} that assumes 
either plane-parallel or spherically symmetric geometry 
\citep{Lester2008}. The intensity profiles are computed for one thousand
equally spaced values of $\mu$, where $\mu \equiv \cos \theta$ and 
$\theta$ is the angle formed between a line-of-sight point on the 
stellar disk and the center of the stellar disk.  The spacing of the 
$\mu$ points represent a significantly higher resolution than previous
models \citep[\textit{e.g.}][]{Wade1985,Claret2000}.  The intensity 
profiles 
are computed for the $B, V, R, I, H, K$, \textit{CoRot} and 
\textit{Kepler} wavebands, and also include best-fit limb-darkening
coefficients for a number of parametrized limb-darkening laws 
\citep{Claret2000,Claret2012}.

The calculation of the planetary transit light curves was done 
using the small-planet approximation of \citet{Mandel2002}.  This 
approximation assumes that the portion of the 
star's surface being blocked by the transiting planet has a uniform
brightness.  Following \cite{Mandel2002}, this approximation is assumed to apply when  
$R_\mathrm{P} \lesssim 0.1 R_\ast$, and is parameterized as
\begin{equation} \label{eq:def_rho} 
\rho \equiv \frac{R_\mathrm{p}}{R_\ast} \lesssim 0.1,
\end{equation}
where $R_\mathrm{P}$ is the radius 
of the planet and $R_\ast$ is the star's radius.  This ratio corresponds approximately to Jupiter transiting the Sun, or to a Uranus-size planet transiting a K dwarf star.  While
this is an approximation adopted to simplify the calculation of the 
light curve, we will show that the differences between our light curves scale approximately as $(R_\mathrm{P}/R_\ast)^2$, and that errors introduced by assuming the small-planet approximation largely cancel out.   For this assumption 
\citet{Mandel2002} derived expressions for the star's relative
flux, $f$, defined as
\begin{equation} \label{eq:rf}
f \equiv \mathrm{\frac{flux \ during \ transit}{unobscured \ flux}}.
\end{equation}
 There are three possible cases depending the projected separation between the
center of the planet and the center of the star normalized by the 
stellar radius, represented as $z$, and the relative sizes of the planet
and star given by $\rho$ defined above.  If the planet does not transit the star, the parameter $z > 1 + \rho$, which clearly gives $f(z) = 1$.
If the planet grazes the stellar disk, $1 - \rho < z \leq 1 + \rho$, 
then
\begin{eqnarray} \label{eq:eq3}
f(z) & = & 1 - \frac{I^\ast(z)}{4 \Omega} 
\left [ \rho^2 \cos^{-1} \left ( \frac{z-1}{\rho} \right ) \right . \nonumber \\
& & \left. - (z-1) \sqrt{\rho^2 - (z-1)^2} \right ],
\end{eqnarray}
where 
\begin{equation} \label{eq:eq4}
I^\ast (z) = (1 - a)^{-1} \int_{z - \rho}^1 2rI(r)dr,
\end{equation}
$a$ is the planet's orbital semimajor axis,
$r \equiv \sin \theta = \sqrt{1 - \mu^2}$, and the term $\Omega$ 
is determined 
by the limb-darkening law assumed.  Using a \citet{Claret2000} 
four-parameter (4-p) law,
\begin{equation} \label{eq:eq5}
\frac{I(\mu)}{I(\mu=1)} = 1 - \sum_{n=1}^4 c_n(1 - \mu^{n/2}),
\end{equation}
gives
\begin{equation} \label{eq:eq6}
\Omega = \sum_{n=0}^4 c_n(n+4)^{-1},
\end{equation}
where $c_0 \equiv 1 - c_1 - c_2 - c_3 - c_4$.  
The third case is for the interior light curve, $z < 1-\rho$, 
\begin{equation} \label{eq:eq7}
f = 1 - \frac{\rho^2 I^\ast(z)}{4\Omega},
\end{equation}
where in this case
\begin{equation} \label{eq:eq8}
I^\ast (z) = (4z \rho)^{-1} \int_{z-\rho}^{z+\rho} 2rI(r)dr.
\end{equation}
The star's intensity distribution contributes to the terms 
$I^\ast(z)$ and $\Omega$ through Equations~\ref{eq:eq4}, \ref{eq:eq6} 
and \ref{eq:eq8}.  The term $\Omega$ was defined by \cite{Mandel2002} based on the four-parameter limb-darkening law, but more generally $\Omega$ is a flux-like variable such that $4\Omega = \int_0^1 2r I( r)dr$.  In the following sections, we compute synthetic planetary transit light curves using these definitions of $4\Omega$ and $I^{*}(z)$.

\section{Center-to-limb intensity variations and limb-darkening laws}\label{s3}
To illustrate the quality of the fits provided by limb-darkening
laws, we use a model stellar atmosphere with the effective temperature $T_{\rm{eff}} = 5600~$K, gravity $\log g = 4.5$ and mass $M_* = 1.1~M_\odot$, from the grid of model stellar atmospheres computed by \cite{Neilson2013b}.  That model assumes solar composition from the \cite{Kurucz2005} opacity files and is computed for both plane-parallel and spherical symmetries.  The plane-parallel geometry assumes that the depth of the atmosphere is small relative to the radius of the star, which simplifies the model radiative transfer greatly as the radiation field depends on depth only \citep{Mihalas1978}.  In that case limb darkening is computed simply as an exponential scaling of the central radiation field.  In spherical symmetry, one relaxes that assumption for a more physically realistic scenario where radiative transfer must be treated as a function of both depth and angle.  As such limb darkening cannot be scaled relative to the central radiation field and must be calculated self-consistently, leading to a different structure as a function of angle, especially near the edge of the star.

The CLIV for each geometry is plotted in Figure~\ref{f1} along with a best-fitting quadratic limb-darkening law of the form
\begin{equation}\label{eq:quad}
\frac{I(\mu)}{I(\mu = 1)} = 1 - u_1(1-\mu) - u_2(1-\mu)^2,
\end{equation}
for the $B,V,R,I,H,K$, \textit{CoRot} and \textit{Kepler} wavebands. This limb-darkening law clearly provides a better fit for the plane-parallel model CLIV than for the spherically symmetric model, but even for the plane-parallel models there is  significant divergence 
near the limb of the star as $\mu \rightarrow 0$. The limb-darkening law deviates more significantly from the model CLIV for the spherically symmetric model.  In terms of wavelength, the difference is greatest towards the near-infrared $H$- and $K$-bands, where  the quadratic law fails to fit the sharp drop in intensity for the spherically symmetric models.  The reason the CLIV has such a sharp drop and step-like structure is related to the total optical depth in the photosphere at infrared wavelengths.  At these longer wavelengths the H$^-$ opacity that dominates the absorption of radiation is smallest and the photosphere is most transparent.  Going from the center of the disk toward the limb the radiation comes from depths of the atmosphere where the temperature declines by only a small amount with increasing height, which leads to a relatively flat CLIV.  However, when $\mu \leq 0.1$, which corresponds to $R \geq 0.99 R_\star$, the atmospheric density becomes so small that the intensity drops rapidly. This indicates that the assumed limb-darkening law introduces errors for fitting light curves, particularly at these near-IR wavelengths.


In Figure~\ref{f2} we repeat the comparison assuming the \cite{Claret2000} 4-p limb-darkening law. The fits of the 4-p law are indistinguishable from the CLIV for the plane-parallel model.  However, the best-fit four-parameter laws still do not fit the limb of the spherically symmetric model CLIV, especially the near-infrared $H$- and $K$-bands. Because the spherically symmetric model stellar atmospheres are more physically representative of actual stellar atmospheres, even the 4-p limb-darkening law will have  deficiencies fitting planetary transit light curves as observations achieve higher precision.


\section{Comparison between CLIV and quadratic limb-darkening laws}\label{s4}
\begin{figure}[th]
\begin{center}
\epsscale{1.15}
\plotone{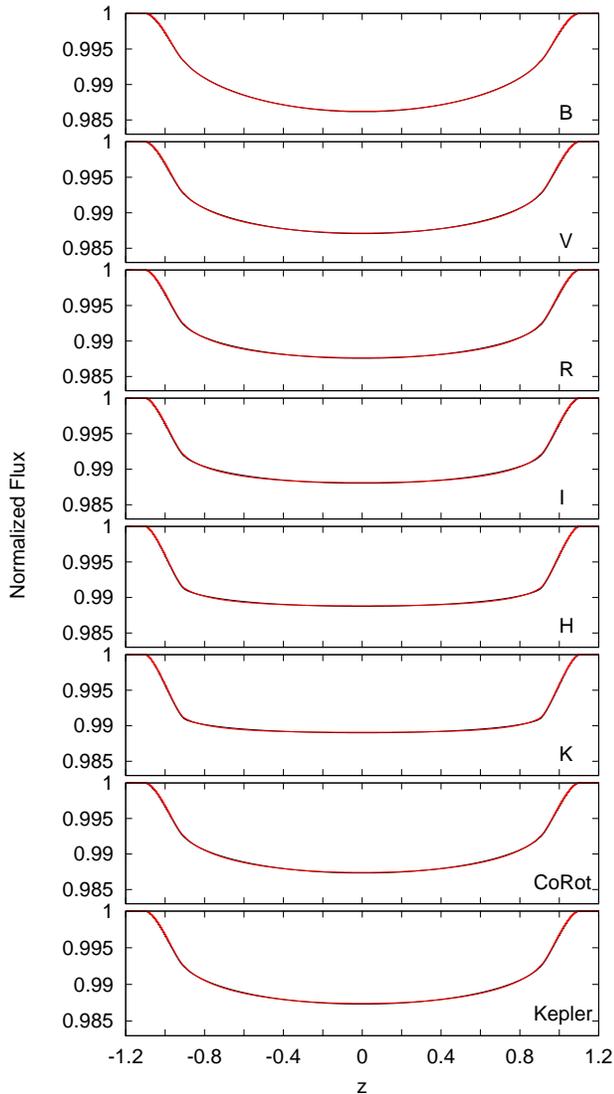}
\end{center}
\caption{The predicted planetary transit light curves  computed using the small-planet approximation, $R_\mathrm{p}/R_* = 0.1$, for a plane-parallel model stellar atmosphere with $T_{\rm{eff}} = 5700~$K and $\log g = 4.5$. There is essentially no difference between the 
transit curves computed using the model CLIV (black) and using the 
best-fit quadratic limb-darkening law (red).}
\label{pp_lc}\end{figure}

The next step is to compare how model CLIV and quadratic limb-darkening laws predict planetary transit light curves using the analytic 
prescription of \cite{Mandel2002}. Their derivation is defined for an edge-on transit,  \textit{i.e.}, an inclination $i = 90^\circ$ and an impact parameter $b = \cos i = 0$. That derivation also assumes the orbit of the transiting planet is circular. 
Light curves were calculated as a function of waveband, and we present the  transit of the orbit in Figure~\ref{pp_lc} for the case of a plane-parallel model stellar atmosphere and corresponding limb-darkening coefficients.



\begin{figure}[t]
\begin{center}
\epsscale{1.15}
\plotone{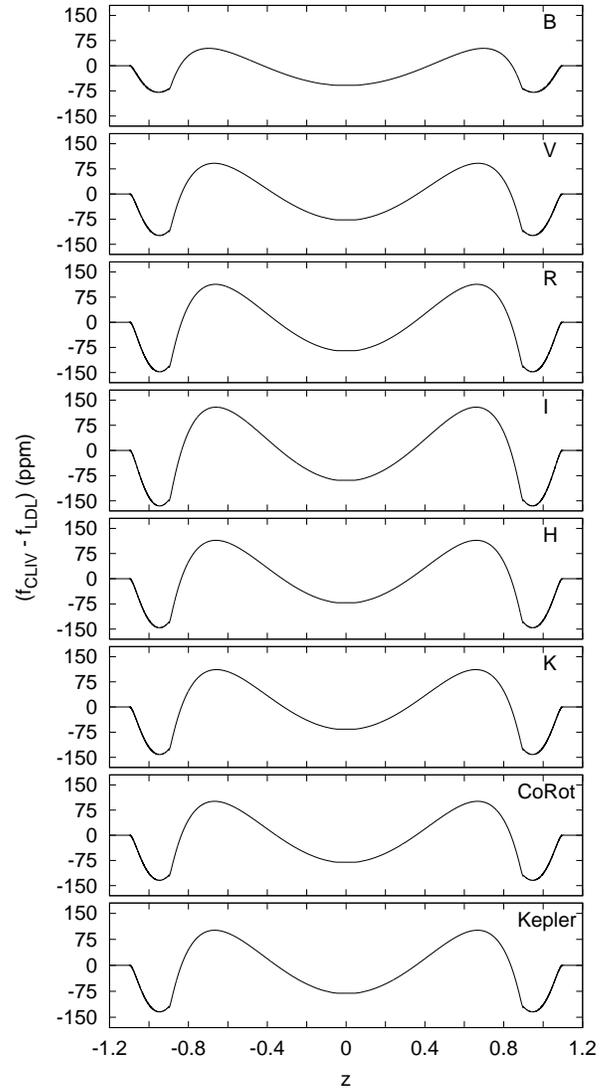}
\end{center}
\caption{The difference between the transit light curves computed from a plane-parallel model stellar atmosphere CLIV and its best-fit quadratic limb-darkening law from Figure~\ref{pp_lc} as a function of wavelength.}
\label{ppdiff}\end{figure}

The results shown in  Figure~\ref{pp_lc} suggest that for 
plane-parallel models there is little difference between using a model CLIV or a limb-darkening law with best-fit coefficients.  However, these plots are somewhat misleading because the plots are dominated by the transit depth, which is controlled by the central part of the stellar disk where the two representations agree.  To distinguish between the two representations, we plot in Figure~\ref{ppdiff} the difference between the light curve derived from the computed
CLIV, $f_\mathrm{CLIV}$, and the light curve derived using 
the limb-darkening law, $f_\mathrm{LDL}$, normalized by the transit depth. Because, by definition, $f \rightarrow 1$ at the stellar limb, the difference will go to zero at the ingress and egress of the planet's transit.    

The difference between the CLIV and the limb-darkening coefficient-derived transits is most important when the planet goes through ingress/egress where the difference is more than about 100~ppm.   This difference is not large, but it is significant and occurs at all wavelengths considered here.  It further suggests that the difference is due solely to the inability of the quadratic limb-darkening law to fit the model CLIV, even for plane-parallel atmospheres.  At $z =  0.9$, the difference in the \textit{Kepler} transit flux is about 100~ ppm, which is similar to the precision required to measure planetary oblateness \citep{Seager2002, Zhu2014}.   While the analysis is qualitative, this difference  suggests that the predictions of \cite{Zhu2014} could be entangled with errors in the assumed limb-darkening law instead of planetary oblateness.  While we cannot confirm this suggestion, it is jarring that the differences between the light curves for a spherical and an oblate planet \citep[][their Figure 1]{Zhu2014} has the similar behavior and magnitude as the differences between plane-parallel light curves using the CLIV and assuming a quadratic limb-darkening law in Figure~\ref{ppdiff}.
\begin{figure}[t]
\begin{center}
\epsscale{1.15}
\plotone{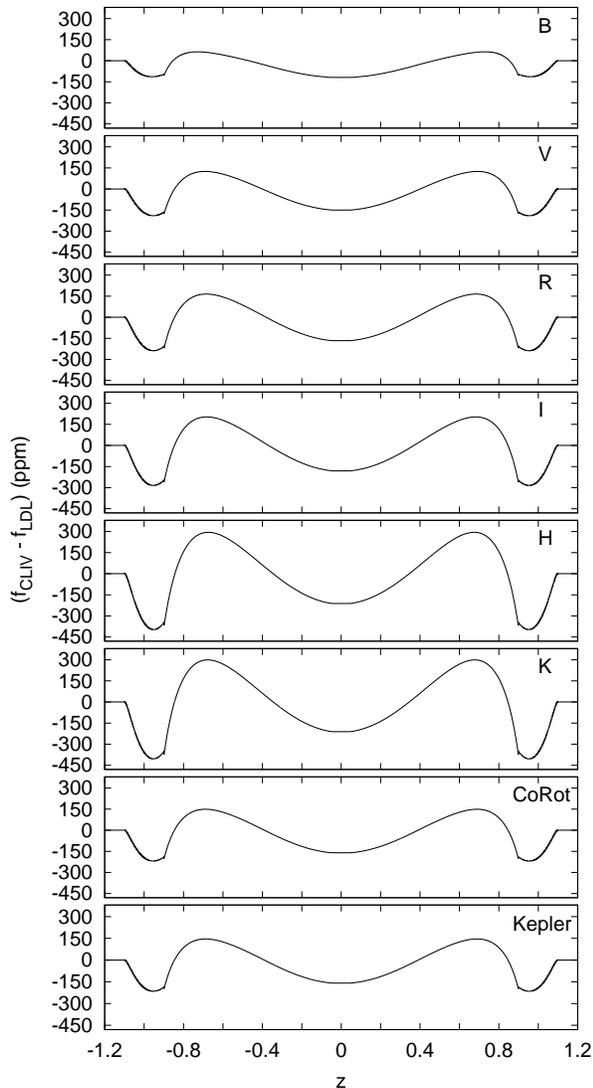}
\end{center}
\caption{The difference between the transit light curves computed from a spherically symmetric model stellar atmosphere CLIV and its best-fit quadratic limb-darkening law as a function of wavelength.}
\label{sphdiff}\end{figure}

We consider how this difference systematically affects the measurement of planetary radii relative to the stellar radii by computing the average difference between the CLIV and limb-darkening law versions of the planetary transit light curves.  This difference is not a measure of the error of the flux introduced by assuming a limb-darkening law.  Instead, the difference provides a  rough estimate of how much the surface area of the planet must change for the light curve computed with the limb-darkening law to agree most closely with light curve computed using the CLIV.  Because this average difference is a measure of the error of the relative surface area of the planet, it tells us about the various errors in transit spectroscopy (which is a differential measure of the surface area as a function of wavelength) and the uncertainty of the planetary radius. For all wavebands, the average difference between the light curve fluxes, hence the relative surface area $(R_\mathrm{p}/R_*)^2$, is about 50~ppm, suggesting an insignificant effect on the best-fit planet radius when we assume plane-parallel model stellar atmospheres.

This first test was for plane-parallel model stellar atmospheres, which have been the conventional approximation for the geometry of a star.  However, a spherically symmetric geometry is a more geometrically realistic representation of a stellar atmosphere, especially for modeling CLIV \citep{Neilson2013a, Neilson2013b}. We repeat our analysis using a spherically symmetric model atmosphere and plot the difference between the planetary transit light curves computed from the CLIV and its best-fit limb-darkening law in Figure~\ref{sphdiff}. The differences are about 3$\times$ greater than in Figure~\ref{ppdiff}, reaching about 400~ppm.  Because differences of this magnitude are now being detected in 
\textit{Kepler} observations, it is  
necessary to fit the observations directly with the CLIV, or at 
least replace the traditional quadratic limb-darkening law with a more 
advanced representation.

%
%
%
Furthermore, for the plane-parallel model the difference at the
transit center between using the CLIV and the best-fit limb-darkening coefficients varies by about 75~ppm at all wavebands.  For the more geometrically realistic spherically symmetric models those errors are greater, with the differences in transit depths varying from about 100~ppm at the shortest wavebands to about 200~ppm for $H$- and $K$-bands and about 150~ppm for the \textit{CoRot} and \textit{Kepler} wavebands. This error is smaller at transit center because, by definition, differences between the CLIV and limb-darkening laws decrease towards the center of the stellar disk.  The differences are due solely to the failure for the limb-darkening law to conserve stellar flux. 

For the spherical models the average difference between the CLIV and the limb-darkening law planetary transit light curves is greater than that computed using plane-parallel models.  At optical wavebands, the average difference is about 60--70~ppm, increasing in the near-IR up to 180~ppm.  This suggests that some of the differences across the transit cancel allowing for a smaller error for measuring the relative planet radius.  For the properties assumed here, the use of limb-darkening laws versus CLIV does not seem to cause large systematic uncertainties.  However, there may be challenges for measuring transit spectroscopy as the differences are a function of wavelength and may inhibit precise measurements of planetary atmospheres and composition.  

The issue of conserving stellar flux from the model stellar atmosphere and the best-fit limb-darkening law is degenerate with the presence of unocculted star spots.  Star spots affect transit light curves by decreasing the stellar flux and making the planet radius seem larger \citep[e.g.][]{Mccullough2014, Hellier2014}.  Because the spots are cooler than the star and follow a blackbody function, the spots  have a differential effect on the stellar flux. As demonstrated in Figures~\ref{ppdiff} and \ref{sphdiff}, the errors in limb darkening representations cause similar errors.  We note, however, that \cite{Mccullough2014} fit starspots that changed the transit depth by a few hundred ppm as opposed to the 100 ppm error we find. 

These errors occur at all wavelengths, but are more significant in the near-infrared $H$- and $K$-bands.  The  differences in the $H$- and $K$-bands are up to about 3\% of the  transit depth due to differences in the curvature of the limb-darkening laws relative to the CLIV. This error may become more important in the era of \textit{WFIRST} and \textit{JWST}, where we expect to observe infrared transits and measure the wavelength sensitivity of the planetary radius.  Because the errors appear to be similar at all wavelengths we concentrate on the \textit{Kepler} optical band and the infrared $K$-band only.

\begin{figure}[t]
\begin{center}
\epsscale{1.15}
\plotone{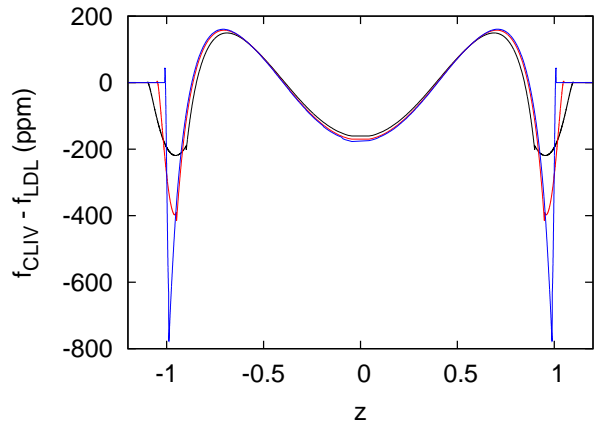}
\end{center}
\caption{Comparison between the differences between planetary transit light curves using CLIV and best-fit quadratic limb-darkening laws for three different planet sizes $R_{\rm{P}}/R_\ast = 0.1$ (black), $0.05$ (red) and $0.01$ (blue) in the {\it Kepler}-waveband.  To make 
the comparisons more comparable, the differences for the smaller planet sizes are scaled by the surface area of the planet relative to the area of the largest planet.}
\label{diff-size}\end{figure}

 We have computed our planetary transits assuming the small 
planet approximation \citep{Mandel2002}, which some may consider too large.  However, we are concerned primarily with the difference between light curves caused by using the 
limb-darkening law so that errors in our calculation due to the small-planet approximation will cancel.  But, we stress that this work is a qualitative comparison to investigate the role of limb-darkening laws in transit observations. We test this by computing test cases with $R_{\rm{P}}/R_\ast = 0.05$ and $0.01$ and plot in Figure~\ref{diff-size} the differences between the  spherically-symmetric model stellar atmosphere CLIV transit light curves and the light curves modeled assuming the quadratic limb-darkening law. To establish a consistent basis of comparison, in this figure we multiply the differences for the $R_{\rm{P}}/R_\ast = 0.05$ and $0.01$ cases by $4$ and $100$, respectively, which are the ratio of surface area blocked by the planet to the star's surface area compared to the blocked surface for the traditional small-planet approximation, $R_{\rm{P}}/R_\ast = 0.1$.  

We see in Figure~\ref{diff-size} that between second and third 
contact of the transit the planet's relative size causes almost no 
change to the differences between the light curves. However, during ingress and egress, the difference as a function of planet size does not scale with surface area simply because only a fraction of planet occults the star, suggesting that as the planet size decreases the relative flux difference increases. This means that using the small-planet approximation underestimates the errors at ingress and egress. But, the comparison is sufficient to show that our analysis is not significantly affected by assuming the traditional small-planet approximation.  The analysis also shows that to first order the error induced by assuming a quadratic limb-darkening law instead of a realistic CLIV scales as a function of the planet's surface area. 

\section{Comparison between CLIV and four-parameter limb-darkening laws}\label{s5}
Section~\ref{s4} explored the quadratic limb-darkening law, which is the most commonly used parameterization for modelling planetary-transit light curves, but \cite{Mandel2002} also derived analytic solutions based on the \cite{Claret2000} four-parameter limb-darkening law, which is considered to be a significant improvement relative to other limb-darkening laws.  \cite{Neilson2013a, Neilson2013b} showed that the four-parameter law was the only law out of six commonly assumed parameterizations tested that could reasonably fit CLIV from spherically symmetric model stellar atmospheres. However, this law requires fitting two more free parameters. We note that the three-parameter law 
of \cite{Sing2009} was not tested, but by its nature the four-parameter limb-darkening law offers a more precise fit to the CLIV, hence we consider that law since it is one of the most precise fits.

\begin{figure*}[t]
\begin{center}
\epsscale{1.15}
\plottwo{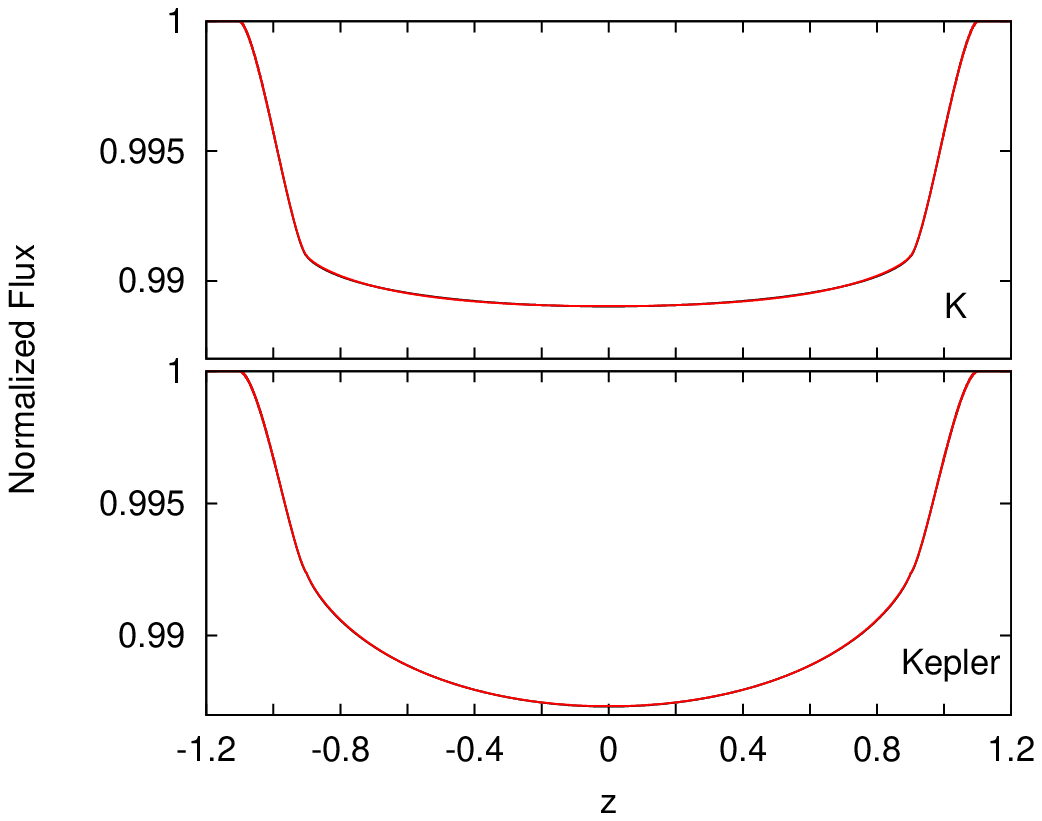}{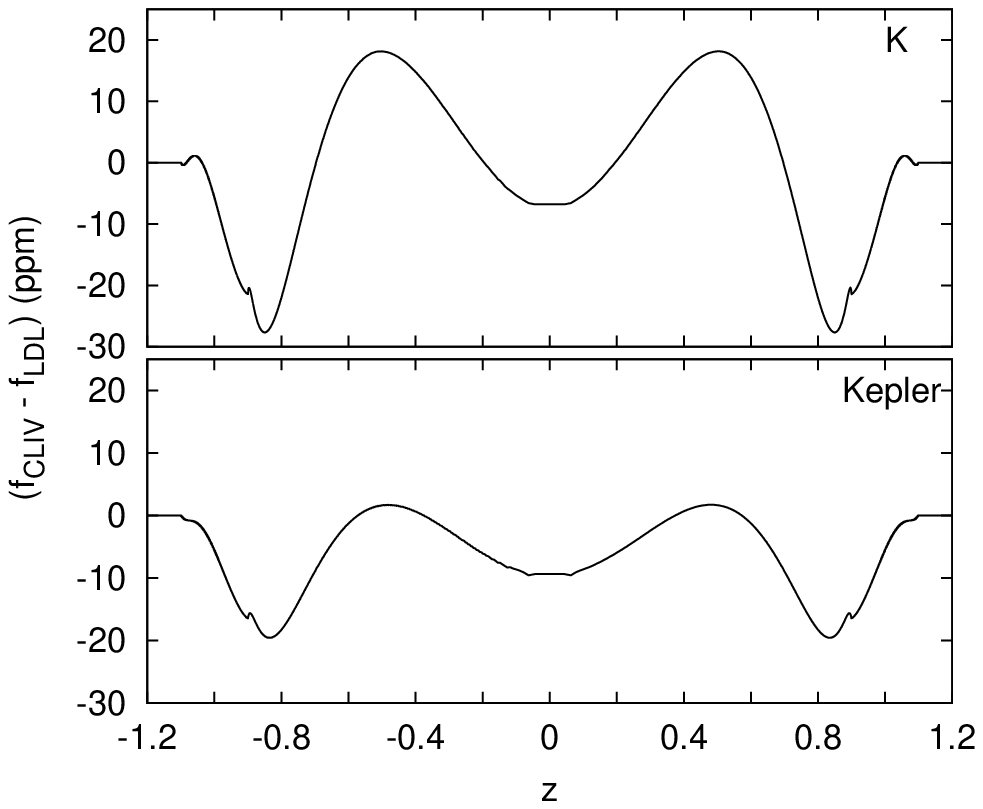}
\end{center}
\caption{(Left) Synthetic planetary-transit light curves for the 
    $K$-band and the \textit{Kepler}-band.  The CLIVs computed using 
    plane-parallel model stellar atmospheres were used directly and to 
    generate the corresponding four-parameter limb-darkening law.
    The two light curves are nearly indistinguishable. (Right) The
    difference between the two light curves  from the Left panel in 
    parts-per-million.}
\label{lp:4p-pp}\end{figure*}

We construct synthetic light curves using the best-fit four-parameter limb-darkening law to compare with those computed directly from the plane-parallel model stellar atmosphere CLIV and present the $K$-band and \textit{Kepler} light curves in Figure~\ref{lp:4p-pp}. There are no noticeable differences between the two light curves at these wavelengths and those presented in previous sections.  To explore this more closely, in the right 
panel of Figure~\ref{lp:4p-pp} we plot the difference between the light curves in parts-per-million.  It is apparent that the addition of two more free parameters to the limb-darkening law only decrease the difference at ingress/egress by a factor of three, leaving a still significant error for the case of plane-parallel models. The agreement is better  at transit center, where the flux difference is about 10~ppm.  While not large, it does suggest a small error for measuring planet size. 

\begin{figure*}[t]
\begin{center}
\epsscale{1.15}
\plottwo{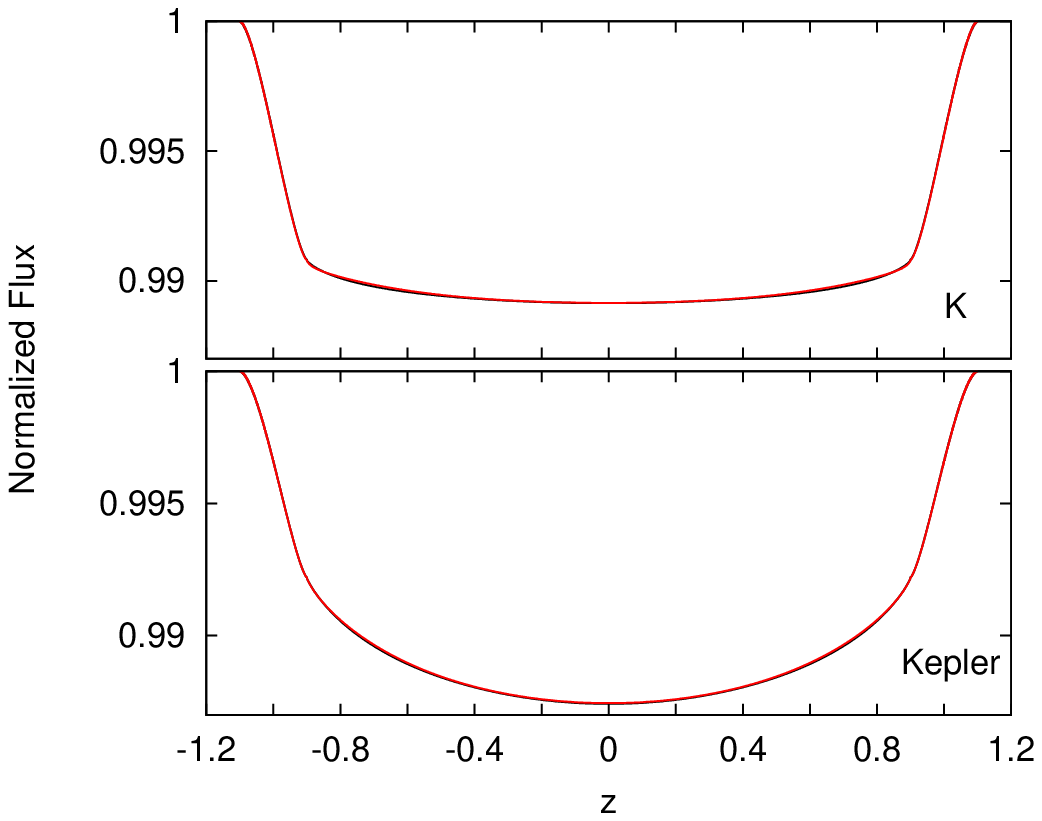}{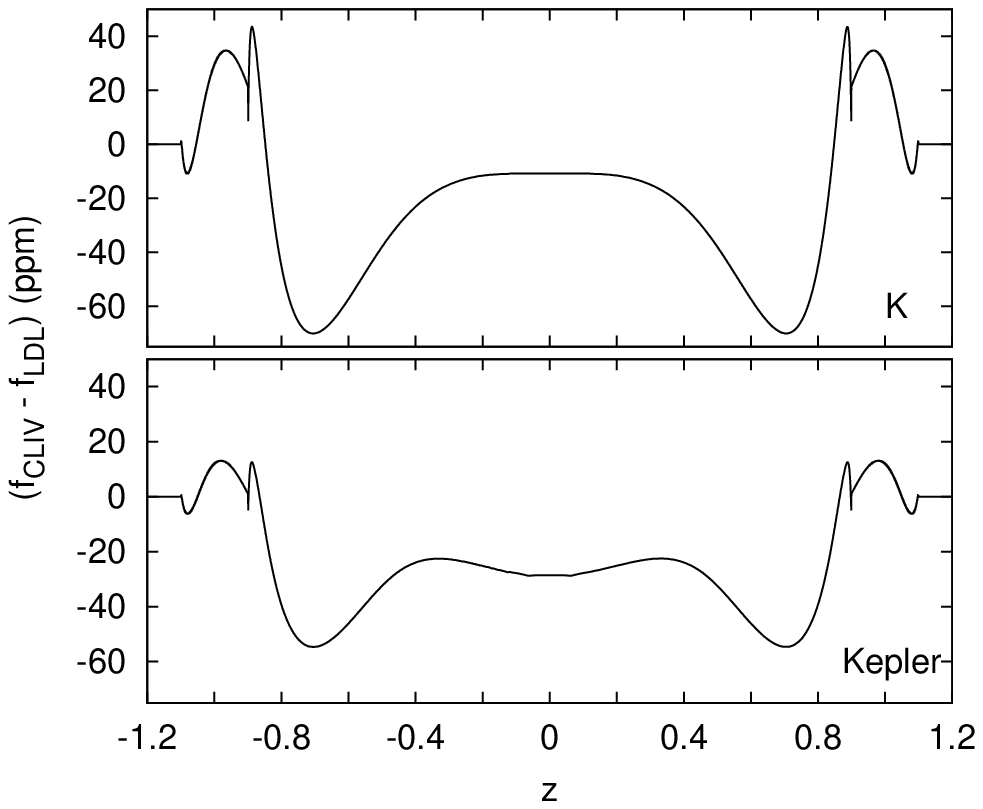}
\end{center}
\caption{(Left) Synthetic planetary-transit light curves for the 
    $K$-band and \textit{Kepler}-band.  The CLIVs computed using 
    spherically symmetric model stellar atmosphere were used directly 
    and to generate the corresponding four-parameter limb-darkening law. 
    The difference between the two light curves are too small to be seen 
    in the plot. (Right) The difference between the light curves in 
    parts-per-million.}
\label{lp:4p-sph}\end{figure*}

Next we repeat the comparison with the more geometrically realistic spherically symmetric model stellar atmospheres. In Figure~\ref{lp:4p-sph}, we show the synthetic light curves and the flux differences.  Again the planet transit light curves are very similar to each other and to the light curves  constructed using plane-parallel models. The differences between the spherically symmetric CLIV and the \cite{Claret2000} 4-p limb-darkening law shown in the right panel are $\leq 40$~ppm for the $K$-band, 
about a factor of 10 improvement compared with the quadratic 
limb-darkening law shown in Figure~\ref{sphdiff}.   This shows 
that the 4-p limb-darkening law is a significant improvement on the quadratic limb-darkening law following the fact that the 4-p law is a more accurate fit to spherically symmetric CLIV.   This shift in limb-darkening laws will be important as we approach the era of 
\textit{TESS}, \textit{PLATO} and \textit{JWST}, but there will still be some small errors.

 It is worth asking how much the relative radius of the planet or the limb-darkening coefficients (LDC) would have to change to minimize the difference between the planetary transit light curves computed using the CLIV and that computed using the LDCs.  One method to test this is to compute a blind Markov Chain Monte Carlo (MCMC) fit of the CLIV light curve that freely fits the LDCs and $\rho$.  However, to perform the MCMC requires injecting noise into the light curve to simulate observational error, which would  introduce assumptions about the observations that would make any results dependent on that noise. Instead we are looking at the best-case scenario that probes how the assumption of limb-darkening laws biases results.   As such, we consider a different approach. 

 We vary the relative radius, $\rho$, until the RMS of the flux differences is minimized while holding the coefficients of the limb-darkening law constant.  We repeat the analysis varying separately each coefficient of the limb-darkening law and using the same relative radius.  To minimize the RMS for the \textit{Kepler}-band we find that $\rho$ must be increased by $\approx 240$-ppm.  Conversely, one could minimize the RMS by decreasing the quadratic coefficient of the limb-darkening law, $u_2$, by about $0.07$ from approximately $u_2 = 0.39$ to $u_2 = 0.32$, but the change in RMS is not significant.  On the other hand, varying the linear term of the limb-darkening law, $u_1$, has little effect on reducing the RMS difference.  Varying either coefficient changes the predicted stellar flux, and hence the transit depth at any point.  However, the shape of the limb-darkening profile is more sensitive to changes of $u_1$ than to changes of $u_2$.  Therefore we see some degeneracy in the \textit{Kepler}-band light curves between the limb-darkening and the radius.  Repeating the experiment for the $K$-band light curves we find the changes must be significantly greater to minimize the RMS differences.  One can either increase $\rho$ by $\approx 450$~ppm or decrease $u_2$ by $\approx 0.2$ from 0.75 to 0.55 to resolve the difference.
  
From this experiment we conclude that if we assume LDCs from model stellar atmospheres that the best-fit planetary radius will have some error simply due to assuming that limb-darkening law.  If we allows the LDCs to vary then we can reduce the remaining error in the radius from $\approx 240$~ppm to $180~$ppm in \emph{Kepler}-band and to about $200~$ppm in the $K$-band.  This suggests that varying the limb-darkening laws will not correct the error significantly. 
 
\section{Significance of the Impact Parameter}\label{s6}

In previous sections we tested how planetary light 
curves vary as a function of the geometry of the model
stellar atmosphere  and the assumed best-fit limb-darkening laws. These 
calculations have all been for planetary transits across the center of the stellar disk, that is for 
an inclination of $90^\circ$ or an impact parameter, $b = 0$. 

 \cite{Howarth2011}  and \cite{Kipping2011a, Kipping2011b} noted that for many inclined extrasolar planet systems, the limb-darkening laws fit directly to observations differ from predicted limb-darkening laws calculated from model stellar atmospheres.  Of course, as noted by \citet{Howarth2011}, the inclination of 
an extrasolar planet's orbit is probably randomly oriented to our line 
of sight, making the limb-darkening fit appropriate for a particular 
inclination different from the limb-darkening fit derived for an 
equatorial transit. This result was especially noted for results presented by \cite{Csizmadia2013} and \cite{Lillo2015}. 

\begin{figure}[t]
\begin{center}
\epsscale{1.15}
\plotone{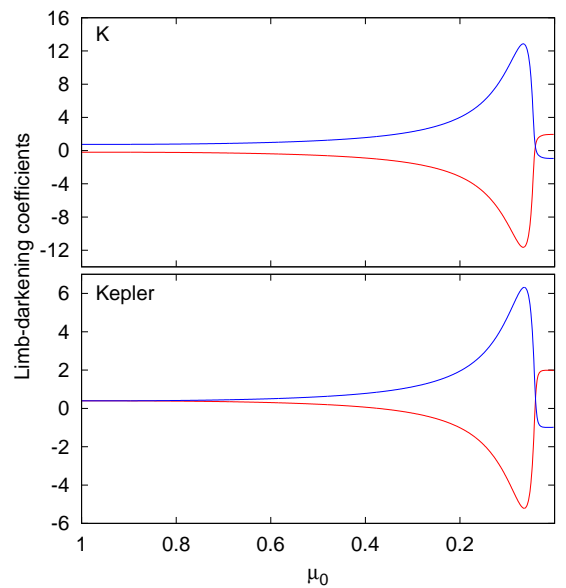}
\end{center}
\caption{The variation of the best-fit limb-darkening coefficients for the quadratic law (equation~\ref{eq:quad}) as a function of impact parameter using the CLIV from the spherically symmetric model stellar atmosphere computed in the previous sections for the $K$-band (top) and \textit{Kepler}-band (bottom).  The red lines denote the linear coefficient, $u_1$, while the blue lines denote the quadratic coefficient, $u_2$.}
\label{limb-darkening coefficients}\end{figure}

In this work, we will adopt a different parameter to represent the inclination of the orbit, 
\begin{equation}\label{eq:mu0}
\mu_0 \equiv \frac{a\cos\theta_0}{R_*} = \frac{a\cos(90^\circ - i)}{R_*}.
\end{equation}
We choose this definition for measuring the inclination to connect the inclination with the definition of $\theta$, the angle between a point on the stellar disk and the center of the star.  As such this definition allows us to write the inclination in terms of the variable $\mu$ to directly explore the connections between the orbital inclination and the CLIV for modeling transits.  
For example, if $b = \cos i = 0.5$ for $a/R_* = 1$, and $\mu_0 = \cos (90^\circ - i) = 0.5$ then the CLIV will be probed from the edge of the disk to an angle $\mu = 0.5$. If the assumed limb-darkening law fits the stellar CLIV accurately then there would be no problem.  However, best-fit limb-darkening coefficients computed from spherically symmetric model stellar atmospheres will be different if we compute them for the whole stellar disk or only part of it.  We show this in Figure~\ref{limb-darkening coefficients} for our computed spherically symmetric \textsc{SAtlas} model stellar atmosphere in the $K$- and \textit{Kepler}-bands.

\begin{figure}[t]
\begin{center}
\epsscale{1.15}
\plotone{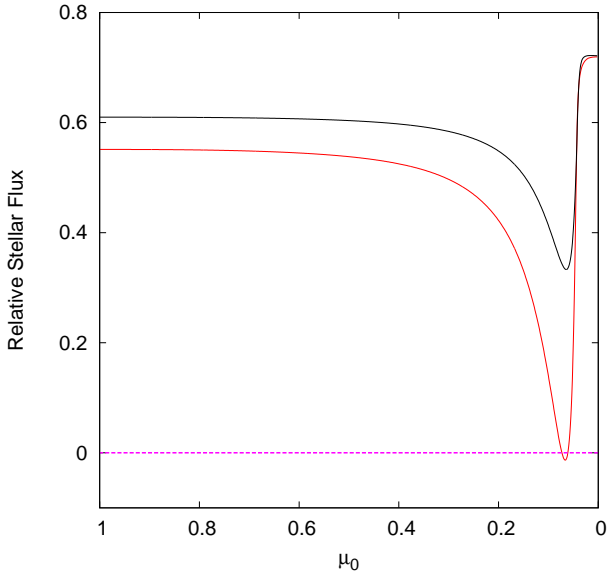}
\end{center}
\caption{Predicted flux normalized with respect to the central intensity computed using best-fit limb-darkening law as a function of impact parameter. The \textit{Kepler}-band flux is shown as the black line and the $K$-band flux is red. The dashed magenta line denotes where the flux is zero. Note that the computed negative values are not physical and are caused solely by the computed limb-darkening coefficients.}
\label{flux}\end{figure}

Figure~\ref{limb-darkening coefficients} shows that there is a large dispersion in the limb-darkening coefficients as a function of impact parameter, especially approaching the limb of the star where the CLIV has the largest gradients as function of $\mu$.  The variation is much greater than presented in the analysis by \cite{Howarth2011} because we are using more geometrically realistic spherically symmetric model stellar atmospheres as opposed to plane-parallel models. This suggests that planetary transit light curves vary significantly as a function of impact parameter and that the  impact parameter influences  the empirical 
determination of the limb-darkening parameters.  It is notable that, for the best-fit coefficients computed here, the predicted stellar flux normalized by the central intensity also varies. The predicted flux from the best-fit limb-darkening laws is shown in Figure~\ref{flux}.  The relative flux becomes slightly negative as $\mu_0 \rightarrow 0$ because the best-fit limb-darkening law cannot accurately map the CLIV of the spherically symmetric atmosphere.

\begin{figure}[t]
\begin{center}
\epsscale{1.15}
\plotone{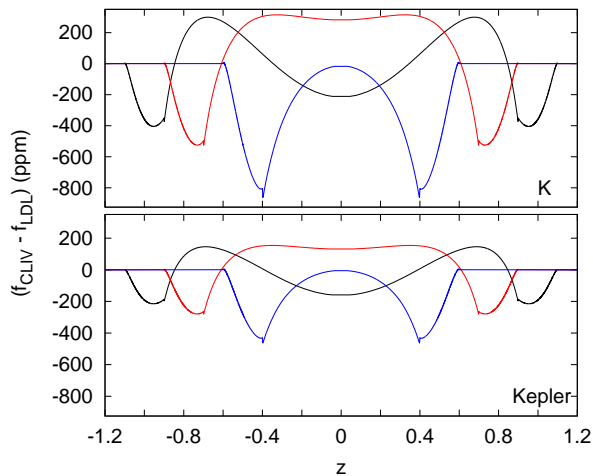}
\end{center}
\caption{Differences between synthetic light curves computed directly from spherically symmetric model stellar atmosphere CLIV and corresponding limb-darkening law for three impact parameters, $\mu_0 = 1$ (black line), $0.8$ (red line) and $0.5$ (blue line). }
\label{transb}\end{figure}

Next we extend our comparison between model stellar atmosphere CLIV and limb-darkening coefficients to investigate impact parameters, $\mu_0 = 1, 0.8, 0.5$.  The limb-darkening coefficients are computed from the entire CLIV as opposed to the approach taken in the hybrid synthetic-photometry/atmospheric-model
(SPAM) method of \cite{Howarth2011}. Figure~\ref{transb} shows the differences between light curves computed directly from the spherical CLIV and those computed from the limb-darkening law in both the $K$- and \textit{Kepler}-bands. As $\mu_0 \rightarrow 0$, the errors grow, supporting the need for changing the limb-darkening coefficients as per the SPAM method.

\begin{figure*}[t]
\begin{center}
\plottwo{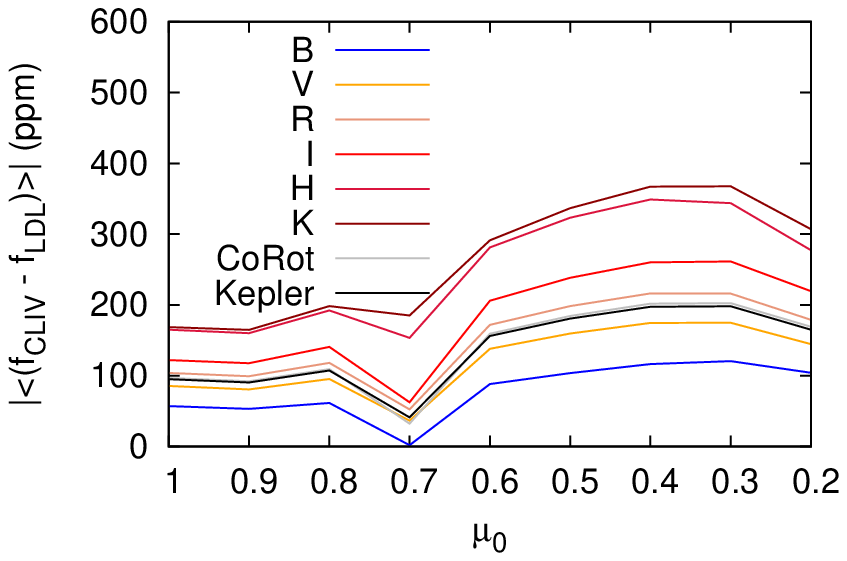}{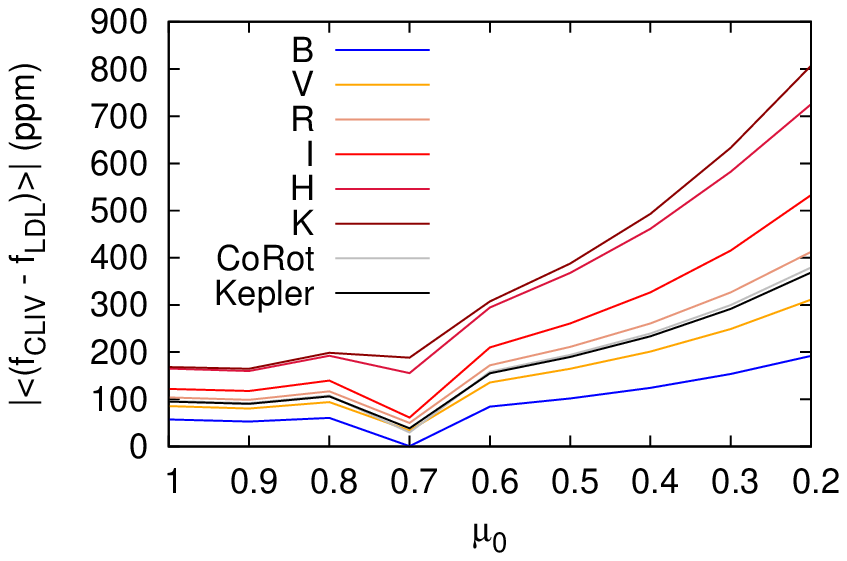}
\end{center}
\caption{The average difference between planetary transit light curves computed using model CLIV and limb-darkening laws as a function of impact parameter. Light curves computed using the same limb-darkening coefficients are shown on the left while those computed using the SPAM method are shown on the right.}
\label{spam}
\end{figure*}

This comparison is generalized in Figure~\ref{spam} to show the difference between the planetary transits computed directly from the spherical model CLIV and transits computed using limb-darkening laws as a function of impact parameter  using the same coefficients for impact parameters (left) and those for the SPAM method (right).  The results suggest that the SPAM method can improve the quality of the fits for impact parameters $0.8 \le \mu_0 \le 0.6$, but gets much worse for smaller impact parameters.

The results shown in Figure~\ref{spam} demonstrate the limitations of the SPAM method and why it may improve best fits to planetary transit observations for only some cases.  The results also imply that for observations of systems that are inclined there will be an extra inherent uncertainty because the best-fit limb-darkening coefficients from observations cannot be both a precise fit of the portion of the CLIV and still conserve stellar flux.  There is a trade off. 

\section{Narrow-Band Spectral Differences}\label{s7}
The analysis presented to this point has concentrated on the differences between broadband spherical model CLIV and best-fit limb-darkening laws. However, it is also important to understand these differences at higher spectral resolution as new observations are exploring spectral properties of extrasolar planets.  From our model stellar atmosphere, we compute CLIV at a nominal spectral resolution of 50 and derived the corresponding best-fit limb-darkening coefficients.  

\begin{figure}[t]
\begin{center}
\plotone{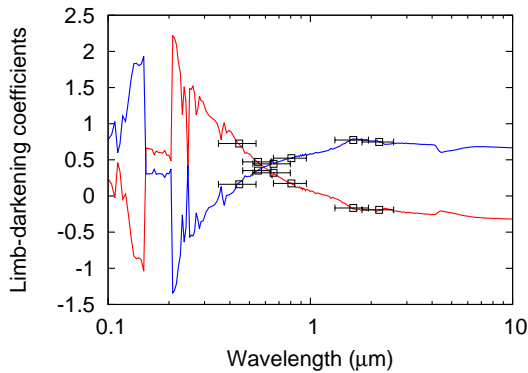}
\end{center}
\caption{Quadratic limb-darkening coefficients as a function of wavelength with a nominal spectral resolving power of 50.  The linear coefficients are shown in red and the quadratic coefficients are in blue.  The limb-darkening coefficients for the broad-band Johnson 
filters are plotted as open squares with error bars indicating the 
bandpass.}
\label{limb-darkening coefficient-w}\end{figure}

The coefficients for the quadratic limb-darkening law are plotted as a function of wavelength in Figure~\ref{limb-darkening coefficient-w} along with those from the broadband coefficients.  At the same wavelength there is close agreement between the broadband coefficients and the coefficients for the higher spectral resolution. However, the higher resolution coefficients vary significantly at other wavelengths,  even including  \emph{limb brightening}.  At longer wavelengths, the coefficients approach a constant value because the CLIV is much flatter and more thermal in nature, being influenced by fewer atomic spectral lines. It should be noted that the limb-darkening coefficients do vary slightly due to molecular opacities.

\begin{figure}[t]
\begin{center}
\plotone{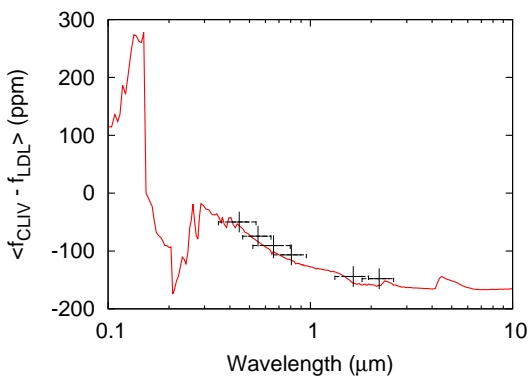}
\end{center}
\caption{The average differences between synthetic light curves computed directly from spherically symmetric model stellar atmosphere CLIV and their corresponding limb-darkening coefficients as a function of wavelength  shown along with the average difference for each Johnson band. }
\label{fig-spectra}\end{figure}

Because of the increasing importance of monochromatic limb darkening and subtle differences with the broadband limb darkening, we compute synthetic planetary transit models as a function of wavelength assuming an impact parameter $\mu_0 = 1$ using our spherically symmetric model stellar atmosphere.  The average difference in flux between the synthetic transits using direct model CLIV and the transits assuming the best-fit limb-darkening coefficients is plotted in Figure~\ref{fig-spectra}. The differences are  similar to the differences for broadband models. 

The difference is greatest in the UV and near UV where the spectrum contains more lines of ions that vary  strongly with depth in the atmosphere.  Because the CLIV probes a range of depths, 
these spectral features prevent a quadratic limb-darkening law from 
achieving a smooth fit.  Of particular interest is that the average difference grows larger, and more negative, going from the optical to the infrared.  This may affect finding evidence of important molecules, such as water, from transit spectral observations that can trace the structure of planetary atmospheres. This difference from $B$- to $K$-band is about 180~ppm, implying that past searches that measure flat planet spectra might be contaminated by the assumption of a quadratic limb-darkening law.  This is because transit spectroscopy is a relative difference in the planet radius (or surface area) as a function of wavelength.  Because the average flux difference in Figure~\ref{fig-spectra} is a measure of the systematic bias of the planetary radius due to the use of this law, it might hide spectral properties of the planet. This would not be an issue for the spectra if the average as a function of wavelength were flat, but we find there is a slope that will contaminate the transit spectra.

These results show that the use of the quadratic limb-darkening law has serious deficiencies, especially towards the infrared wavelengths where the CLIV is more ``box"-like.  \cite{Kreidberg2014} found that infrared planetary transit observations of GJ~1214b could be well-modeled using linear limb-darkening laws.  However, the linear law is a reasonable representation of the infrared CLIV only over a restricted range from the center of the stellar disk because the intensity decreases dramatically near the limb.  This decrease would occur very close to the edge of the disk for a star with the higher gravity of GJ~1214 \citep{Neilson2013b} and appears to be consistent with the residuals measured by \cite{Kreidberg2014}. We will explore the differences between computing transit light curves assuming CLIV and quadratic limb-darkening laws in a forthcoming article.

These differences as a function of wavelength, which range from about $-50$~ppm in the $B$-band to about $-150$~ppm in the infrared, could be significant enough to hide features in measured planetary spectra from transit observations.  For instance,  \cite{Brogi2015} presented measurements of the composition and rotation of HD~189733b using high-resolution infrared spectra, but the spectral features are of the same order of magnitude as the differences presented in Figure~\ref{fig-spectra}.  Similar issues are apparent in the transit spectra analyzed by \cite{Madhusudhan2014}, who also included star spots to fit the observations. This is especially concerning as our results are for low-resolution spectra while they were using high-resolution spectra for which stellar limb darkening can be more complex.

\section{Conclusions}\label{s8}
In this work, we have demonstrated the  importance of stellar CLIV compared to  assumed limb-darkening laws for modeling planetary transit measurements, at least in a qualitative sense.  Specifically, we find that synthetic transit light curves computed directly from spherically symmetric model stellar atmosphere CLIV differ significantly from  light curves computed assuming traditional quadratic limb-darkening laws. For broadband measurements, these differences are up to about 100~to 400~ppm at the limb of the star and the average differences range from about 50 to about 150~ppm.  We note that changing from the small-planet approximation to a more realistic method will not significantly change these results. These results also highlight the challenges for measuring high-precision exoplanet properties when assuming simple, parametric limb-darkening laws.

The average difference measured is small but systematic, and it increases toward the infrared, potentially impacting the uncertainty of spectro-photometric transit measurements of extrasolar planet atmospheres. We tested this importance  by computing the average difference between light curves computed directly from CLIV and from best-fit limb-darkening laws as a function of wavelength.  

We  explored the potential of the SPAM method \citep{Howarth2011} for modeling planetary transits with different impact parameters. While it is certainly essential to take the impact parameter into 
account, we find a flaw in the SPAM method;  by fitting the limb-darkening coefficients to only the part of the star's CLIV  corresponding to the path of the transiting planet,  the  stellar flux is not measured accurately.  As such, the SPAM method is a trade off: improved modeling of the shape of the limb-darkening profile over the region of interest while decreasing the  precision in conserving the stellar flux.  One way around this is to use both sets of limb-darkening coefficients: the SPAM coefficients and the  coefficients fit to the entire disk to measure the transit.  

Our findings suggest that assuming a simple limb-darkening law leads to an intrinsic error for measuring the relative radius of an extrasolar planet and the secondary properties, such as the planet's atmospheres and oblateness. This error is currently small, $\le 200$~ppm, but as observations push to higher precision and new missions such as \textit{TESS}, \textit{PLATO}, and \textit{JWST} start returning observations these differences will become more important.  

We recommend  the following changes to how extrasolar planet transits are modeled:
\begin{enumerate}
\item Directly fit the CLIV  from spherical model stellar atmospheres to the observations, or
\item Switch to the \cite{Claret2000} four-parameter limb-darkening law fit to the CLIV of spherical model stellar atmospheres as a more precise representation of the CLIV, or
\item After fitting observations using a simple quadratic limb-darkening law, compare that law with spherical model stellar atmosphere CLIV \citep[e.g.,][]{Neilson2013b} to quantify the uncertainty introduced.
\end{enumerate}
With regards to our third recommendation, we will present broadband errors introduced by assuming the quadratic limb-darkening law as a function of stellar properties and orbital inclination in a future paper.  However, our first recommendation is our preferred approach.
 In this case, we shift from fitting limb-darkening coefficients to fitting stellar properties such as effective temperature, gravity and stellar mass, which offers a way to understand both the planet and its star to a new precision, especially when coupled with other methods such as the flicker-gravity relation \citep{Bastien2014, Bastien2015}. 

\acknowledgements Ignace is grateful for funding from the NSF (AST-0807664). McNeil acknowledges funding from the Honors College at East Tennessee State University. Lester is grateful for support from a Discovery grant from the Natural Sciences and Engineering Research Council of 
Canada.
\bibliographystyle{aa}
\bibliography{planets}

\begin{thebibliography}{51}
\expandafter\ifx\csname natexlab\endcsname\relax\def\natexlab#1{#1}\fi

\bibitem[{{Bastien} {et~al.}(2016){Bastien}, {Stassun}, {Basri}, \&
  {Pepper}}]{Bastien2015}
{Bastien}, F.~A., {Stassun}, K.~G., {Basri}, G., \& {Pepper}, J. 2016, \apj,
  818, 43

\bibitem[{{Bastien} {et~al.}(2014){Bastien}, {Stassun}, \&
  {Pepper}}]{Bastien2014}
{Bastien}, F.~A., {Stassun}, K.~G., \& {Pepper}, J. 2014, \apjl, 788, L9

\bibitem[{{Borucki} {et~al.}(2010){Borucki}, {Koch}, {Basri}, {Batalha},
  {Brown}, {Caldwell}, {Caldwell}, {Christensen-Dalsgaard}, {Cochran},
  {DeVore}, {Dunham}, {Dupree}, {Gautier}, {Geary}, {Gilliland}, {Gould},
  {Howell}, {Jenkins}, {Kondo}, {Latham}, {Marcy}, {Meibom}, {Kjeldsen},
  {Lissauer}, {Monet}, {Morrison}, {Sasselov}, {Tarter}, {Boss}, {Brownlee},
  {Owen}, {Buzasi}, {Charbonneau}, {Doyle}, {Fortney}, {Ford}, {Holman},
  {Seager}, {Steffen}, {Welsh}, {Rowe}, {Anderson}, {Buchhave}, {Ciardi},
  {Walkowicz}, {Sherry}, {Horch}, {Isaacson}, {Everett}, {Fischer}, {Torres},
  {Johnson}, {Endl}, {MacQueen}, {Bryson}, {Dotson}, {Haas}, {Kolodziejczak},
  {Van Cleve}, {Chandrasekaran}, {Twicken}, {Quintana}, {Clarke}, {Allen},
  {Li}, {Wu}, {Tenenbaum}, {Verner}, {Bruhweiler}, {Barnes}, \&
  {Prsa}}]{Borucki2010}
{Borucki}, W.~J., {Koch}, D., {Basri}, G., {et~al.} 2010, Science, 327, 977

\bibitem[{{Brogi} {et~al.}(2016){Brogi}, {de Kok}, {Albrecht}, {Snellen},
  {Birkby}, \& {Schwarz}}]{Brogi2015}
{Brogi}, M., {de Kok}, R.~J., {Albrecht}, S., {et~al.} 2016, \apj, 817, 106

\bibitem[{{Cabrera} {et~al.}(2010){Cabrera}, {Bruntt}, {Ollivier},
  {D{\'{\i}}az}, {Csizmadia}, {Aigrain}, {Alonso}, {Almenara}, {Auvergne},
  {Baglin}, {Barge}, {Bonomo}, {Bord{\'e}}, {Bouchy}, {Carone}, {Carpano},
  {Deleuil}, {Deeg}, {Dvorak}, {Erikson}, {Ferraz-Mello}, {Fridlund},
  {Gandolfi}, {Gazzano}, {Gillon}, {Guenther}, {Guillot}, {Hatzes}, {Havel},
  {H{\'e}brard}, {Jorda}, {L{\'e}ger}, {Llebaria}, {Lammer}, {Lovis}, {Mazeh},
  {Moutou}, {Ofir}, {von Paris}, {P{\"a}tzold}, {Queloz}, {Rauer}, {Rouan},
  {Santerne}, {Schneider}, {Tingley}, {Titz-Weider}, \&
  {Wuchterl}}]{Cabrera2010}
{Cabrera}, J., {Bruntt}, H., {Ollivier}, M., {et~al.} 2010, \aap, 522, A110

\bibitem[{{Charbonneau} {et~al.}(2000){Charbonneau}, {Brown}, {Latham}, \&
  {Mayor}}]{Charbonneau2004}
{Charbonneau}, D., {Brown}, T.~M., {Latham}, D.~W., \& {Mayor}, M. 2000, \apjl,
  529, L45

\bibitem[{{Chiavassa} {et~al.}(2010){Chiavassa}, {Collet}, {Casagrande}, \&
  {Asplund}}]{Chiavassa2010}
{Chiavassa}, A., {Collet}, R., {Casagrande}, L., \& {Asplund}, M. 2010, \aap,
  524, A93

\bibitem[{{Claret}(2000)}]{Claret2000}
{Claret}, A. 2000, \aap, 363, 1081

\bibitem[{{Claret}(2009)}]{Claret2009}
{Claret}, A. 2009, \aap, 506, 1335

\bibitem[{{Claret} {et~al.}(2012){Claret}, {Hauschildt}, \&
  {Witte}}]{Claret2012}
{Claret}, A., {Hauschildt}, P.~H., \& {Witte}, S. 2012, \aap, 546, A14

\bibitem[{{Csizmadia} {et~al.}(2013){Csizmadia}, {Pasternacki}, {Dreyer},
  {Cabrera}, {Erikson}, \& {Rauer}}]{Csizmadia2013}
{Csizmadia}, S., {Pasternacki}, T., {Dreyer}, C., {et~al.} 2013, \aap, 549, A9

\bibitem[{{Deming} {et~al.}(2011){Deming}, {Sada}, {Jackson}, {Peterson},
  {Agol}, {Knutson}, {Jennings}, {Haase}, \& {Bays}}]{Deming2011}
{Deming}, D., {Sada}, P.~V., {Jackson}, B., {et~al.} 2011, \apj, 740, 33

\bibitem[{{Dominik}(2004)}]{Dominik2004}
{Dominik}, M. 2004, \mnras, 352, 1315

\bibitem[{{Espinoza} \& {Jord{\'a}n}(2015)}]{Espinoza2015}
{Espinoza}, N. \& {Jord{\'a}n}, A. 2015, \mnras, 450, 1879

\bibitem[{{Espinoza} \& {Jord{\'a}n}(2016)}]{Espinoza2016}
{Espinoza}, N. \& {Jord{\'a}n}, A. 2016, \mnras, 457, 3573

\bibitem[{{Fouqu{\'e}} {et~al.}(2010){Fouqu{\'e}}, {Heyrovsk{\'y}}, {Dong},
  {Gould}, {Udalski}, {Albrow}, {Batista}, {Beaulieu}, {Bennett}, {Bond},
  {Bramich}, {Calchi Novati}, {Cassan}, {Coutures}, {Dieters}, {Dominik},
  {Dominis Prester}, {Greenhill}, {Horne}, {J{\o}rgensen}, {Koz{\l}owski},
  {Kubas}, {Lee}, {Marquette}, {Mathiasen}, {Menzies}, {Monard}, {Nishiyama},
  {Papadakis}, {Street}, {Sumi}, {Williams}, {Yee}, {Brillant}, {Caldwell},
  {Cole}, {Cook}, {Donatowicz}, {Kains}, {Kane}, {Martin}, {Pollard}, {Sahu},
  {Tsapras}, {Wambsganss}, {Depoy}, {Gaudi}, {Han}, {Lee}, {Park}, {Kubiak},
  {Szyma{\'n}ski}, {Pietrzy{\'n}ski}, {Soszy{\'n}ski}, {Szewczyk}, {Ulaczyk},
  {Abe}, {Fukui}, {Furusawa}, {Gilmore}, {Hearnshaw}, {Itow}, {Kamiya},
  {Kilmartin}, {Korpela}, {Lin}, {Ling}, {Masuda}, {Matsubara}, {Miyake},
  {Muraki}, {Nagaya}, {Ohnishi}, {Okumura}, {Perrott}, {Rattenbury}, {Saito},
  {Sako}, {Sato}, {Skuljan}, {Sullivan}, {Sweatman}, {Tristram}, {Allan},
  {Bode}, {Burgdorf}, {Clay}, {Fraser}, {Hawkins}, {Kerins}, {Lister},
  {Mottram}, {Saunders}, {Snodgrass}, {Steele}, {Anguita}, {Bozza},
  {Harps{\o}e}, {Hinse}, {Hundertmark}, {Kj{\ae}rgaard}, {Liebig}, {Mancini},
  {Masi}, {Rahvar}, {Ricci}, {Scarpetta}, {Southworth}, {Surdej}, {Th{\"o}ne},
  {Riffeser}, \& {Seitz}}]{Fouque2011}
{Fouqu{\'e}}, P., {Heyrovsk{\'y}}, D., {Dong}, S., {et~al.} 2010, \aap, 518,
  A51

\bibitem[{{Hellier} {et~al.}(2014){Hellier}, {Anderson}, {Cameron}, {Delrez},
  {Gillon}, {Jehin}, {Lendl}, {Maxted}, {Pepe}, {Pollacco}, {Queloz},
  {S{\'e}gransan}, {Smalley}, {Smith}, {Southworth}, {Triaud}, {Udry}, \&
  {West}}]{Hellier2014}
{Hellier}, C., {Anderson}, D.~R., {Cameron}, A.~C., {et~al.} 2014, \mnras, 440,
  1982

\bibitem[{{Howarth}(2011)}]{Howarth2011}
{Howarth}, I.~D. 2011, \mnras, 418, 1165

\bibitem[{{Kipping} \& {Bakos}(2011{\natexlab{a}})}]{Kipping2011a}
{Kipping}, D. \& {Bakos}, G. 2011{\natexlab{a}}, \apj, 730, 50

\bibitem[{{Kipping} \& {Bakos}(2011{\natexlab{b}})}]{Kipping2011b}
{Kipping}, D. \& {Bakos}, G. 2011{\natexlab{b}}, \apj, 733, 36

\bibitem[{{Kipping}(2013)}]{Kipping2013}
{Kipping}, D.~M. 2013, \mnras, 435, 2152

\bibitem[{{Kirk} {et~al.}(2016){Kirk}, {Conroy}, {Pr{\v s}a}, {Abdul-Masih},
  {Kochoska}, {Matijevi{\v c}}, {Hambleton}, {Barclay}, {Bloemen}, {Boyajian},
  {Doyle}, {Fulton}, {Hoekstra}, {Jek}, {Kane}, {Kostov}, {Latham}, {Mazeh},
  {Orosz}, {Pepper}, {Quarles}, {Ragozzine}, {Shporer}, {Southworth},
  {Stassun}, {Thompson}, {Welsh}, {Agol}, {Derekas}, {Devor}, {Fischer},
  {Green}, {Gropp}, {Jacobs}, {Johnston}, {LaCourse}, {Saetre}, {Schwengeler},
  {Toczyski}, {Werner}, {Garrett}, {Gore}, {Martinez}, {Spitzer}, {Stevick},
  {Thomadis}, {Halley Vrijmoet}, {Yenawine}, {Batalha}, \&
  {Borucki}}]{Kirk2016}
{Kirk}, B., {Conroy}, K., {Pr{\v s}a}, A., {et~al.} 2016, \aj, 151, 68

\bibitem[{{Knutson} {et~al.}(2007){Knutson}, {Charbonneau}, {Noyes}, {Brown},
  \& {Gilliland}}]{Knutson2007}
{Knutson}, H.~A., {Charbonneau}, D., {Noyes}, R.~W., {Brown}, T.~M., \&
  {Gilliland}, R.~L. 2007, \apj, 655, 564

\bibitem[{{Koch} {et~al.}(2010){Koch}, {Borucki}, {Basri}, {Batalha}, {Brown},
  {Caldwell}, {Christensen-Dalsgaard}, {Cochran}, {DeVore}, {Dunham},
  {Gautier}, {Geary}, {Gilliland}, {Gould}, {Jenkins}, {Kondo}, {Latham},
  {Lissauer}, {Marcy}, {Monet}, {Sasselov}, {Boss}, {Brownlee}, {Caldwell},
  {Dupree}, {Howell}, {Kjeldsen}, {Meibom}, {Morrison}, {Owen}, {Reitsema},
  {Tarter}, {Bryson}, {Dotson}, {Gazis}, {Haas}, {Kolodziejczak}, {Rowe}, {Van
  Cleve}, {Allen}, {Chandrasekaran}, {Clarke}, {Li}, {Quintana}, {Tenenbaum},
  {Twicken}, \& {Wu}}]{Koch2010}
{Koch}, D.~G., {Borucki}, W.~J., {Basri}, G., {et~al.} 2010, \apjl, 713, L79

\bibitem[{{Kreidberg} {et~al.}(2014){Kreidberg}, {Bean}, {D{\'e}sert},
  {Benneke}, {Deming}, {Stevenson}, {Seager}, {Berta-Thompson}, {Seifahrt}, \&
  {Homeier}}]{Kreidberg2014}
{Kreidberg}, L., {Bean}, J.~L., {D{\'e}sert}, J.-M., {et~al.} 2014, \nat, 505,
  69

\bibitem[{{Kurucz}(1979)}]{Kurucz1979}
{Kurucz}, R.~L. 1979, \apjs, 40, 1

\bibitem[{{Kurucz}(2005)}]{Kurucz2005}
{Kurucz}, R.~L. 2005, Memorie della Societa Astronomica Italiana Supplementi,
  8, 14

\bibitem[{{Lester} \& {Neilson}(2008)}]{Lester2008}
{Lester}, J.~B. \& {Neilson}, H.~R. 2008, \aap, 491, 633

\bibitem[{{Lillo-Box} {et~al.}(2015){Lillo-Box}, {Barrado}, {Santos},
  {Mancini}, {Figueira}, {Ciceri}, \& {Henning}}]{Lillo2015}
{Lillo-Box}, J., {Barrado}, D., {Santos}, N.~C., {et~al.} 2015, \aap, 577, A105

\bibitem[{{Madhusudhan} {et~al.}(2014){Madhusudhan}, {Crouzet}, {McCullough},
  {Deming}, \& {Hedges}}]{Madhusudhan2014}
{Madhusudhan}, N., {Crouzet}, N., {McCullough}, P.~R., {Deming}, D., \&
  {Hedges}, C. 2014, \apjl, 791, L9

\bibitem[{{Mandel} \& {Agol}(2002)}]{Mandel2002}
{Mandel}, K. \& {Agol}, E. 2002, \apjl, 580, L171

\bibitem[{{McCullough} {et~al.}(2014){McCullough}, {Crouzet}, {Deming}, \&
  {Madhusudhan}}]{Mccullough2014}
{McCullough}, P.~R., {Crouzet}, N., {Deming}, D., \& {Madhusudhan}, N. 2014,
  \apj, 791, 55

\bibitem[{{Mihalas}(1978)}]{Mihalas1978}
{Mihalas}, D. 1978, {Stellar atmospheres /2nd edition/}

\bibitem[{{M{\"u}ller} {et~al.}(2013){M{\"u}ller}, {Huber}, {Czesla}, {Wolter},
  \& {Schmitt}}]{Muller2013}
{M{\"u}ller}, H.~M., {Huber}, K.~F., {Czesla}, S., {Wolter}, U., \& {Schmitt},
  J.~H.~M.~M. 2013, \aap, 560, A112

\bibitem[{{Neilson} \& {Lester}(2011)}]{Neilson2011}
{Neilson}, H.~R. \& {Lester}, J.~B. 2011, \aap, 530, A65

\bibitem[{{Neilson} \& {Lester}(2012)}]{Neilson2012}
{Neilson}, H.~R. \& {Lester}, J.~B. 2012, \aap, 544, A117

\bibitem[{{Neilson} \& {Lester}(2013{\natexlab{a}})}]{Neilson2013a}
{Neilson}, H.~R. \& {Lester}, J.~B. 2013{\natexlab{a}}, \aap, 554, A98

\bibitem[{{Neilson} \& {Lester}(2013{\natexlab{b}})}]{Neilson2013b}
{Neilson}, H.~R. \& {Lester}, J.~B. 2013{\natexlab{b}}, \aap, 556, A86

\bibitem[{{Pollacco} {et~al.}(2006){Pollacco}, {Skillen}, {Collier Cameron},
  {Christian}, {Hellier}, {Irwin}, {Lister}, {Street}, {West}, {Anderson},
  {Clarkson}, {Deeg}, {Enoch}, {Evans}, {Fitzsimmons}, {Haswell}, {Hodgkin},
  {Horne}, {Kane}, {Keenan}, {Maxted}, {Norton}, {Osborne}, {Parley}, {Ryans},
  {Smalley}, {Wheatley}, \& {Wilson}}]{Pollacco2006}
{Pollacco}, D.~L., {Skillen}, I., {Collier Cameron}, A., {et~al.} 2006, \pasp,
  118, 1407

\bibitem[{{Popper}(1984)}]{Popper1984}
{Popper}, D.~M. 1984, \aj, 89, 132

\bibitem[{{Pr{\v s}a} \& {Zwitter}(2005)}]{Prsa2005}
{Pr{\v s}a}, A. \& {Zwitter}, T. 2005, \apj, 628, 426

\bibitem[{{Seager} \& {Hui}(2002)}]{Seager2002}
{Seager}, S. \& {Hui}, L. 2002, \apj, 574, 1004

\bibitem[{{Sing}(2010)}]{Sing2010}
{Sing}, D.~K. 2010, \aap, 510, A21

\bibitem[{{Sing} {et~al.}(2009){Sing}, {D{\'e}sert}, {Lecavelier Des Etangs},
  {Ballester}, {Vidal-Madjar}, {Parmentier}, {Hebrard}, \& {Henry}}]{Sing2009}
{Sing}, D.~K., {D{\'e}sert}, J.-M., {Lecavelier Des Etangs}, A., {et~al.} 2009,
  \aap, 505, 891

\bibitem[{{Tsiaras} {et~al.}(2016){Tsiaras}, {Rocchetto}, {Waldmann}, {Venot},
  {Varley}, {Morello}, {Damiano}, {Tinetti}, {Barton}, {Yurchenko}, \&
  {Tennyson}}]{Tsiaras2015}
{Tsiaras}, A., {Rocchetto}, M., {Waldmann}, I.~P., {et~al.} 2016, \apj, 820, 99

\bibitem[{{Wade} \& {Rucinski}(1985)}]{Wade1985}
{Wade}, R.~A. \& {Rucinski}, S.~M. 1985, \aaps, 60, 471

\bibitem[{{Winn} {et~al.}(2007){Winn}, {Holman}, {Bakos}, {P{\'a}l}, {Johnson},
  {Williams}, {Shporer}, {Mazeh}, {Fernandez}, {Latham}, \&
  {Gillon}}]{Winn2007}
{Winn}, J.~N., {Holman}, M.~J., {Bakos}, G.~{\'A}., {et~al.} 2007, \aj, 134,
  1707

\bibitem[{{Wittkowski} {et~al.}(2006{\natexlab{a}}){Wittkowski}, {Aufdenberg},
  {Driebe}, {Roccatagliata}, {Szeifert}, \& {Wolff}}]{Wittkowski2006b}
{Wittkowski}, M., {Aufdenberg}, J.~P., {Driebe}, T., {et~al.}
  2006{\natexlab{a}}, \aap, 460, 855

\bibitem[{{Wittkowski} {et~al.}(2004){Wittkowski}, {Aufdenberg}, \&
  {Kervella}}]{Wittkowski2004}
{Wittkowski}, M., {Aufdenberg}, J.~P., \& {Kervella}, P. 2004, \aap, 413, 711

\bibitem[{{Wittkowski} {et~al.}(2006{\natexlab{b}}){Wittkowski}, {Hummel},
  {Aufdenberg}, \& {Roccatagliata}}]{Wittkowski2006a}
{Wittkowski}, M., {Hummel}, C.~A., {Aufdenberg}, J.~P., \& {Roccatagliata}, V.
  2006{\natexlab{b}}, \aap, 460, 843

\bibitem[{{Zhu} {et~al.}(2014){Zhu}, {Huang}, {Zhou}, \& {Lin}}]{Zhu2014}
{Zhu}, W., {Huang}, C.~X., {Zhou}, G., \& {Lin}, D.~N.~C. 2014, \apj, 796, 67

\end{thebibliography}

\end{document}